# Potential Dependent Ionic Sieving Through Functionalized Laminar MoS$_2$ Membranes

*Wisit Hirunpinyopas,[1,2] Eric Prestat,[3,4] Pawin Iamprasertkun,[1,2] Mark A. Bissett,[1,3]\* Robert A. W. Dryfe[1,2]\**

[1]National Graphene Institute, [2]School of Chemistry, and [3]School of Materials, University of Manchester, Oxford Road, Manchester, M13 9PL, United Kingdom
[4]SuperSTEM Laboratory, SciTech Daresbury Campus, Daresbury, WA4 4AD, United Kingdom
E-mail: mark.bissett@manchester.ac.uk or robert.dryfe@manchester.ac.uk*

**Abstract**

Laminar MoS$_2$ membranes show outstanding potential for practical applications in energy conversion/storage, sensing, and as nanofluidic devices. For water purification technologies, MoS$_2$ membranes can form abundant nanocapillaries from layered stacks of exfoliated MoS$_2$ nanosheets. These MoS$_2$ membranes have previously demonstrated excellent ionic rejection with high water permeation rates, as well as long-term stability with no significant swelling when exposed to aqueous or organic solvents. Chemical modification of these MoS$_2$ membranes has been shown to improve their ionic rejection properties, however the mechanism behind this improvement is not well understood. To elucidate this mechanism we report the potential dependant ion transport through functionalized MoS$_2$ membranes. The ionic permeability of the MoS$_2$ membrane was transformed by chemical functionalization with a simple naphthalene sulfonate dye (sunset yellow) and found to decrease by over a factor of ~10 compared to the pristine MoS$_2$ membranes and those reported for graphene oxide and Ti$_3$C$_2$T$_x$ (MXene) membranes. The effect of pH, solute concentration, and ionic size/charge on the ionic selectivity of the functionalized MoS$_2$ membranes is also reported. The potential dependant study of these dye functionalized MoS$_2$ membranes for ionic sieving with charge selectivity should enable future applications in electro-dialysis and ion exchange for water treatment technologies.

1. **Introduction**

Two dimensional (2D) materials, in the form of laminar membranes, have been widely studied for water purification applications such as desalination, ion exchange, and electro-dialysis.[1-3] Recently, graphene and graphene oxide (GO), when formed into layers of randomly re-stacked nanosheets, have shown to be promising membrane materials, demonstrating high rejection properties towards small ionic solutes while maintaining high water permeation rates due to a network of nanocapillary channels formed between the individual layered materials.[3-6] However, GO-based membranes are reported to be unstable when submerged in aqueous solutions as the membranes become swollen. These membranes consequently have an enlarged interlayer spacing resulting in poor rejection properties.[7, 8] To overcome this, a number of membrane modification strategies have been developed, including additives to crosslink individual layers[8, 9] and physical confinement[3] to restrict the membrane's swelling. Membranes modified by these approaches exhibited improved rejection properties but a decrease in the water flux. This drawback will hinder capacity for large scale exploitation in industry.

Other 2D materials have also been reported as suitable to form laminar membranes, e.g. transition metal carbides/nitrides, MXenes (e.g. $Ti_3C_2$ and $Ti_3CN$),[10, 11] and transition metal dichalcogenides, TMDs (e.g. $MoS_2$ and $WS_2$).[12-14] Deng et al.[12] reported that $MoS_2$ laminar membranes showed good performance with high stability under extreme pH conditions, without any significant expansion of the interlayer distance. These promising $MoS_2$ membranes are not only stable in a range of aqueous media but also show significantly higher water permeation rates than similar laminar membranes. Despite this promise there are few studies on the ion transport through $MoS_2$ membranes; however recent reports have demonstrated improved ionic rejection, surpassing similar GO membranes, by chemical functionalization but the mechanism for this improvement is unclear. TMD laminar membranes also demonstrated molecular sieving/separation performance for both organic vapour and liquid media which lends them to extensive uses in membrane technology applications under extreme conditions.[12, 14]

Chemical functionalization is crucial in altering the desired surface chemistry of 2D materials, providing control over the membrane's properties for use in various applications such as water treatment,[1, 15] energy storage,[16] and photocatalysis[17]. For water purification, we have previously described the functionalization of $MoS_2$ membranes with organic dyes, which exhibit an increase in water flux by a factor of 4, with high ion rejection (~99%) when compared to membranes of pristine, exfoliated $MoS_2$ of comparable thickness.[1] These dye functionalized $MoS_2$ membranes also showed ca. 5-fold increase in water flux compared to the previously described GO-based membranes.[3, 4]

Ion transport through GO-based membranes under the influence of an applied potential has been studied previously by Hong et al.[18], where the membranes were found to exhibit high ionic rejection resulting from the electrostatic repulsion and the size exclusion imparted by the negative surface charge and confinement by the nanochannels. These GO membranes were, however, of extremely limited applicability due to their aforementioned tendency to swell in aqueous solutions, leading to poor ionic rejection. Specifically, the interlayer spacing increases to nearly 70 Å when the membrane was completely soaked in deionized water.[7]

In this work, we firstly investigated ion mobility through laminar $MoS_2$ membranes in the presence of an applied potential and salt concentration gradient. As-prepared functionalized membranes are stable in a range of aqueous solutions including acidic and basic media (no delamination was detected). The sunset yellow (disodium 6-hydroxy-5-[(4-sulfophenyl)azo]-2-naphthalenesulfonate; anionic dye) functionalized, laminar $MoS_2$ membranes (MoS$_2$/SY) show significant retardation of ion transport as well as capability for size and charge selectivity compared to the pristine MoS$_2$ membranes, previously reported laminar membranes (GO and Ti$_3$C$_2$T$_x$), and commercial polymeric membranes (Nafion membranes). A range of characterization techniques including optical microscopy, electron microscopy (scanning electron microscopy, SEM, and scanning transmission electron microscopy, STEM), powder X-ray diffraction (PXRD), X-ray photoelectron spectroscopy (XPS), and zeta ($\zeta$) potential measurements were employed to determine the quality of the MoS$_2$ membrane, along with its thickness, stability, chemical composition, and surface charge. Moreover, MoS$_2$/SY membranes were also characterized as a function of permeate concentration and solution pH, thereby demonstrating their excellent cation selectivity at low permeate concentration and high pH condition.

## 2. Results & Discussion

The ionic mobility of dye functionalized MoS$_2$ membranes was measured as schematically shown in Figure 1a; after first using the functionalization procedure described in our previous work (see Figure S1-2).[1] Briefly, 1 mM sunset yellow (SY) was used to functionalize the laminar MoS$_2$ membranes and they were then cleaned to remove any excess dye molecules until no remaining dye was detected, by UV-visible spectroscopy and electro-spray mass spectroscopy. The characterization of the pristine and dye functionalized MoS$_2$ membranes are provided as Supporting Information. Figure 1b shows the experimental setup for ion transport in a variety of salt solutions, under a concentration gradient, with 100 mM and 10 mM in the feed and permeate reservoirs, respectively, using a four-electrode system.[19-21] The MoS$_2$ membranes were supported on polyvinylidene fluoride (PVDF) membranes during their pressure filtration driven by self-assembly as shown in Figure 1c. Figure 1d shows a cross-sectional SEM image of a MoS$_2$/PVDF membrane showing the laminar structure of the MoS$_2$ flakes, with a lateral size range of 200-300 nm. The high resolution plan view and cross-sectional SEMs are shown in Figure S3-4, as well as energy dispersive X-ray (EDX) analysis mapping of Mo and S elements in Figure S4. Figure 1e reveals the laminar structure of the stacked MoS$_2$ flakes using scanning transmission electron microscopy (STEM). The individual MoS$_2$ flakes are composed of ca. ~1-15 layers, with the majority being 3-5 layers thick (Figure S5). The individual sheets are predominantly stacked horizontally, with some vertically-aligned, randomly oriented flakes also present. The channel height measured between individual MoS$_2$ flakes is approximately 6-13 Å. The PXRD pattern of MoS$_2$/SY also indicates the membrane stability (no significant swelling) after long exposure in aqueous solution as seen by the consistency of the (002) peak position as shown in Figure S6.

The assembled MoS$_2$ membrane was inserted in a custom-made H-beaker cell consisting of two liquid reservoirs (50 mL) contacted with Pt electrodes and Ag/AgCl reference electrodes, connected to the main solution via 3M KCl agarose salt bridges inside Luggin capillaries (placed ~2 mm on either side of the membrane to minimize the membranes' resistance) as shown in Figure 1b. Pt and Ag/AgCl electrodes in the feed chamber were connected to the counter and reference electrodes terminals of the potentiostat, whereas the corresponding electrodes in the permeate chamber were connected to the working and sense terminals. The transmembrane potential was cycled using a triangular potential waveform from −200 mV to 200 mV at a rate of 1 mV s$^{-1}$ with a reverse cycle to ensure there was no hysteresis in the response. Figure 1f shows the *I-V* characteristics of different valence ions, measured at a constant concentration ratio (10 mM/100 mM). As a function of the size and charge of the ions, the membrane potential (zero-current potential, the applied potential at which there is no net flow of ions) decreased to a more negative potential with increasing cationic charge. The net current at zero applied voltage is indicative of different diffusion rates between cations and anions resulting in the shift of the *I-V* response along the voltage axis for both the functionalized (Figure 1f) and pristine $MoS_2$ membranes (Figure S7). A negative current at zero applied voltage corresponds to the higher mobility of cations compared to that of $Cl^-$ anions, vice versa for a positive current as shown in Figure 1f. The permeation properties of the $MoS_2$ membrane are therefore transformed by dye functionalization. The effect of SY functionalization is clearly seen in Figure 1g, which reveals that the membrane conductance decreased significantly, with a related fall in the zero-current potential resulting in a 2-fold reduction of mobility ratio as determined by the Goldman-Hodgkin-Katz (GHK) equation (Eq. 1), compared to the pristine $MoS_2$. The conductance of salt with three different cationic charges (KCl, $BaCl_2$, and $AlCl_3$) has been plotted as a function of $MoS_2$/SY thickness as shown in Figure 1h. This analysis shows that the trivalent salt with larger hydrated cation radius ($AlCl_3$) has the highest conductance for a given membrane thickness. Moreover, the conductance values decreased dramatically with increasing membrane thickness, from 1 μm to 3 μm, but were only slightly lower for thicker (5 μm) membranes. This can be explained in terms of either an imperfect packing of the laminar $MoS_2$ membranes, as supported by the STEM image of the laminar $MoS_2$ membrane (Figure 1e), or non-uniform dye functionalization of the thicker membrane.[1] This reduction in the effect of increasing membrane thickness may be attributed to the increased dye functionalization at the outer surface of the membrane, with limited dye diffusion through the inner $MoS_2$ membrane.

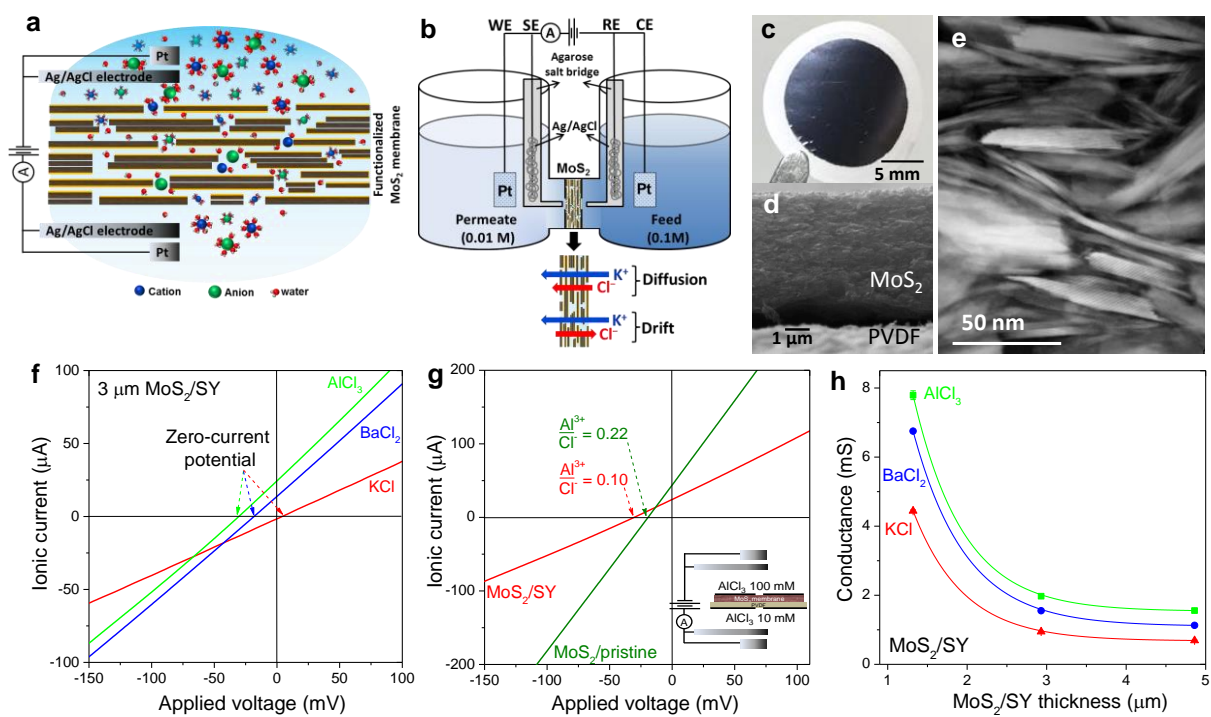

**Figure 1.** Ion transport through nanocapillary channels. (a) Schematic showing ion transport through a sunset yellow (SY) functionalized laminar $MoS_2$ membrane ($MoS_2$/SY). (b) Schematic of the experimental setup used to investigate ion transport using a four-electrode system, as well as a schematic showing a drift-diffusion experiment. Ag/AgCl electrodes contact the main solution via 3M KCl agarose salt bridges to eliminate the liquid-junction potentials. (c) Photograph of the $MoS_2$ membranes supported on a PVDF membrane. (d) Cross-sectional SEM image of the $MoS_2$ on PVDF support demonstrating the laminar $MoS_2$ structure. $MoS_2$ membranes were produced with three different thicknesses (1, 3, and 5 μm). (e) Cross-sectional high angle annular dark field (HAADF) STEM image of the pristine $MoS_2$ membrane supported on the PVDF. (f) I–V characteristics of $MoS_2$/SY (3.01 ± 0.13 μm thick), showing three different ionic salts measured under at a fixed concentration ratio (10 mM/100 mM). (g) I–V characteristics of $AlCl_3$ for $MoS_2$/SY and pristine $MoS_2$ with comparable thickness (~3 μm thick), with the $Al^{3+}/Cl^-$ mobility ratio calculated by the GHK current equation (Eq.1) for those membranes. (h) Conductance ($G = I/\Delta V$) of the $MoS_2$/SY membrane for different valence cations (KCl, $BaCl_2$, and $AlCl_3$) as a function of $MoS_2$/SY thickness. All ionic conductances reported had the conductance of a bare PVDF membrane subtracted from the measured value (see Supporting Information).

To quantify the influence of charge on ion transport through the MoS$_2$ membranes, *I-V* measurements known as 'drift-diffusion' experiments were performed, as they are driven by both the diffusion due to concentration gradient and the applied voltage difference.[18, 22, 23] The GHK current equation, which assumes independence of the ion movements across a membrane, was used to express the cation/anion mobility ratio ($\mu^+/\mu^-$) as shown in Eq. 1.[18, 24-26]

$$\frac{\mu^+}{\mu^-} = -\frac{z_-^2}{z_+^2}\left(\frac{[C^-]_f - [C^-]_p \exp(z_- \frac{FE_m}{RT})}{[C^+]_f - [C^+]_p \exp(z_+ \frac{FE_m}{RT})}\right)\left(\frac{1 - \exp(z_+ \frac{FE_m}{RT})}{1 - \exp(z_- \frac{FE_m}{RT})}\right) \quad (1)$$

where $E_m$ is the membrane potential (zero-current potential, often called the reversal potential), $[C^-]_f$ and $[C^+]_p$ are the concentration of anions and cations in the feed and the permeate reservoirs, respectively, $z_+$ and $z_-$ are the valences of cations and anions, respectively, and other symbols have their usual meanings. The mobility ratio, $\mu^+/\mu^-$, of the pristine exfoliated and functionalized MoS$_2$ membranes were plotted using Eq. 1 as shown in Figure 2a. The mobility ratios of the pristine and functionalized MoS$_2$ membranes decreased with increasing hydrated cation radii, which changed over one order of magnitude from K$^+$ to Al$^{3+}$ ions, after the treatment of the chemical functionalization. This is in agreement with previous work studying ion transport through angstrom-scale slits.[25] Moreover, the individual mobility of cations ($\mu^+$) and anions ($\mu^-$) were calculated from the relation between ionic conductivity ($\sigma$) and ion mobility:

$$\sigma \approx F(C^+\mu^+ + C^-\mu^-) \quad (2)$$

The conductivity of chloride solutions was measured using a relatively high concentration of 0.1 M in both feed and permeate reservoirs (see Table S1-2 for a comparison of the ionic conductivity between the MoS$_2$ membranes and bulk chloride solutions), which allows the surface charge contribution to be neglected.[18, 25, 27] Figure 2b plots the ion mobility (●■ symbols) of the cations, and their corresponding chloride counter ions (○□ symbols) within the $MoS_2$ membranes, as a function of hydrated cation radii[25, 28, 29] obtained by the combination of Eq. 2 and the calculated mobility ratio. The cation mobility ($\mu^+$) in laminar $MoS_2$ membranes decreased by ca. 10 and 100-fold for a pristine $MoS_2$ membrane and the dye functionalized $MoS_2$ membrane, respectively, compared to ion mobility in bulk solutions (♦◊ symbols).[30] Moreover, our measured ion mobilities are also compared to the literature values reported for the same type of laminar membranes (graphene oxide (see Supporting Information: Figure S8) and $Ti_3C_2T_x$)[11, 18] and porous polymeric separation membranes (Nafion-117 and Nafion XL perfluorosulfonic acid (PFSA) membranes)[31, 32] as shown in Figure 2b. This analysis indicated that a pristine $MoS_2$ membrane exhibited the retardation of ion transport of a similar magnitude to previously reported membranes. The variation of $Cl^-$ mobility was found to be within 15% and 20% for the pristine and the dye functionalization, respectively, estimated from average $Cl^-$ mobility for several ionic salt solutions. This finding is in good agreement with the variation in $Cl^-$ mobility found for transport through Ångström-scale slits (±15%).[25] The cation mobility decreased with increasing hydrated cation radius, $R_H$, from $K^+$ to $Al^{3+}$ with the trend lines shown across cation radii, especially for $MoS_2$/SY. The dye functionalized $MoS_2$ membrane not only significantly suppressed ion transport through the membranes but also decreased the mobility for the cations with the larger hydrated radii. This is clearly revealed by comparing cation mobility between the functionalized and pristine $MoS_2$ membranes, which correspond to ca. 9.5, 10.3, and 17.9-fold reduction in cation mobility for a range of mono-, di-, and trivalent cations, respectively, for the dye functionalized $MoS_2$ membrane compared to its pristine counterpart. By comparing the similarly sized ions $K^+$ (3.31 Å) and $Cl^-$ (3.32 Å),[28] the potassium cations are transported through $MoS_2$ membranes slightly more quickly than the chloride anion ($K^+/Cl^-$ >

1), an effect attributed to the negatively charged surface of MoS$_2$ membranes. This was observed for both pristine and dye functionalized MoS$_2$ membranes.

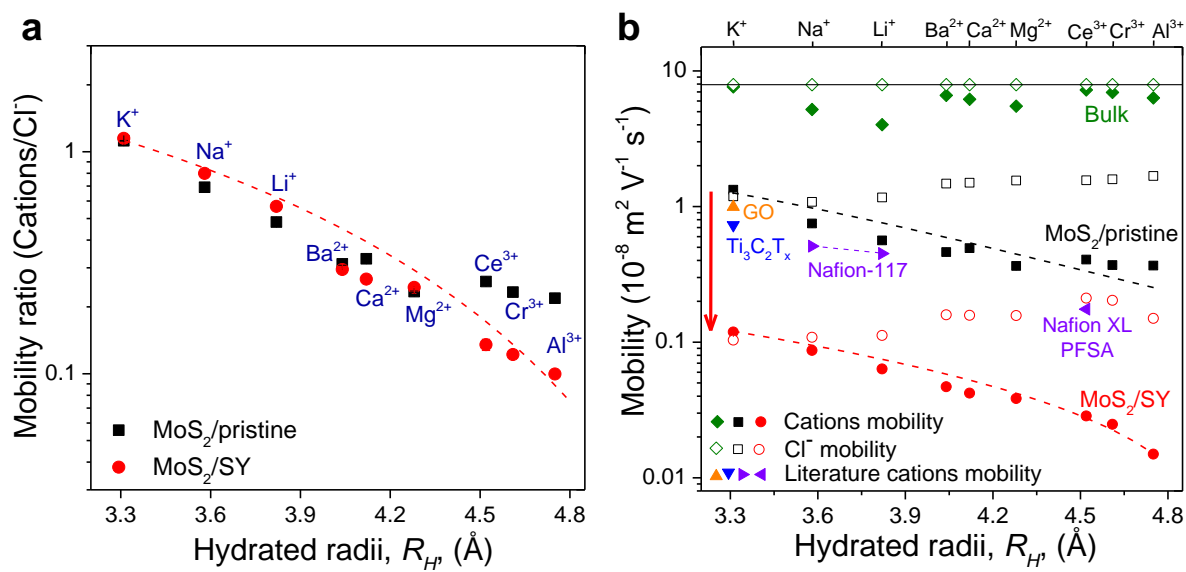

**Figure 2.** Ion mobility through the MoS$_2$ membranes. (a) Mobility ratio, $\mu^+/\mu^-$, as a function of hydrated cation radius ($R_H$) for 3.01 ± 0.13 μm thick MoS$_2$/SY (● symbols) and 2.93 ± 0.08 μm thick pristine MoS$_2$ (■ symbols). (b) Ion mobility of cations (■● symbols) as well as their chloride counter ions (□○ symbols) of the MoS$_2$ membranes as a function of $R_H$. Diamonds (♦◇ symbols) represent literature values for cations and anions in bulk solutions,[30] while all triangles (▲, ▼, ▶, and ◀ symbols) represent the cation mobility reported through GO-based membranes (see Supporting Information),[18] Ti$_3$C$_2$T$_x$ (MXene) membranes,[11] Nafion-117 membranes,[31] and Nafion XL PFSA membranes,[32] respectively. The ion mobility ($\mu_i$) reported in GO and Nafion-117 membranes were calculated from ion diffusion coefficients ($D_i$) using the Nernst-Einstein relation ($D_i = \mu_i RT/F$). Curves in (a) and (b) added are guides to the eye. Error bars for (a) and (b) are negligible on this scale. A red arrow in (b) indicates the reduction of mobility for MoS$_2$/SY compared to MoS$_2$/pristine. Note the semi-logarithmic scale.

Furthermore, the transport through the MoS$_2$ membranes was evaluated with different concentration of the permeate reservoirs from $10^{-2}$ M to $10^{-5}$ M using KCl solution with a constant feed concentration at 0.1 M: the resultant current-voltage ($I$–$V$) responses are given in Figure 3a. By decreasing the KCl concentration on the permeate side, the diffusion current at zero applied voltage rapidly dropped due to the suppressed transport of Cl⁻ through the MoS$_2$/SY membrane as shown in Figure 3b. To understand the effect of diffusive pressure on the ion transport across the MoS$_2$ membrane, we measured the zero-current potential, as well as the Cl⁻ mobility, for a variety of KCl permeate molarities as shown in Figure 3c. The Cl⁻ mobility exponentially dropped through the MoS$_2$ membrane with the decreasing salt concentration at the permeate side in agreement with previous literature using graphene oxide membranes,[18] single-layer MoS$_2$/graphene nanopores,[22, 23] carbon/boron nitride nanotube,[33, 34] and nanochannel slits.[25, 27, 35] This is due to the effective surface charge of the membrane, which dominates membrane transport at the low salt concentration in the permeate reservoir.[35, 36] This suggestion is also supported by the zeta potential measurements showing that the functionalized MoS$_2$ nanosheets are negatively charged at pH values around 7. The zero-current potentials increased in proportion to the log of the permeate molarity, which results in the increase of the calculated K⁺/Cl⁻ mobility ratio (K⁺ >> Cl⁻). In addition, the ionic selectivity (%$S$) at different permeate molarity can be deduced from the ionic mobility ($\mu_i$) defined as;

$$\%S_+ = \left(\frac{\mu^+}{\mu^+ + \mu^-}\right) \times 100 \quad (3)$$

Figure 3d shows ion selectivity as a function of KCl concentration in the permeate reservoir. The selectivity for both ions (K⁺ and Cl⁻) are nearly similar (~50%) for the highest concentration ratio (10 mM/100 mM), and diverges with decreasing KCl concentration at the permeate side. This phenomenon can be explained by the increase of the surface zeta potential, due to the decrease of ionic strength as previously reported in pristine MoS$_2$ membranes.[37] The increasing zeta potentials result in repulsion between negatively charged MoS$_2$/SY surface and Cl⁻ ions.

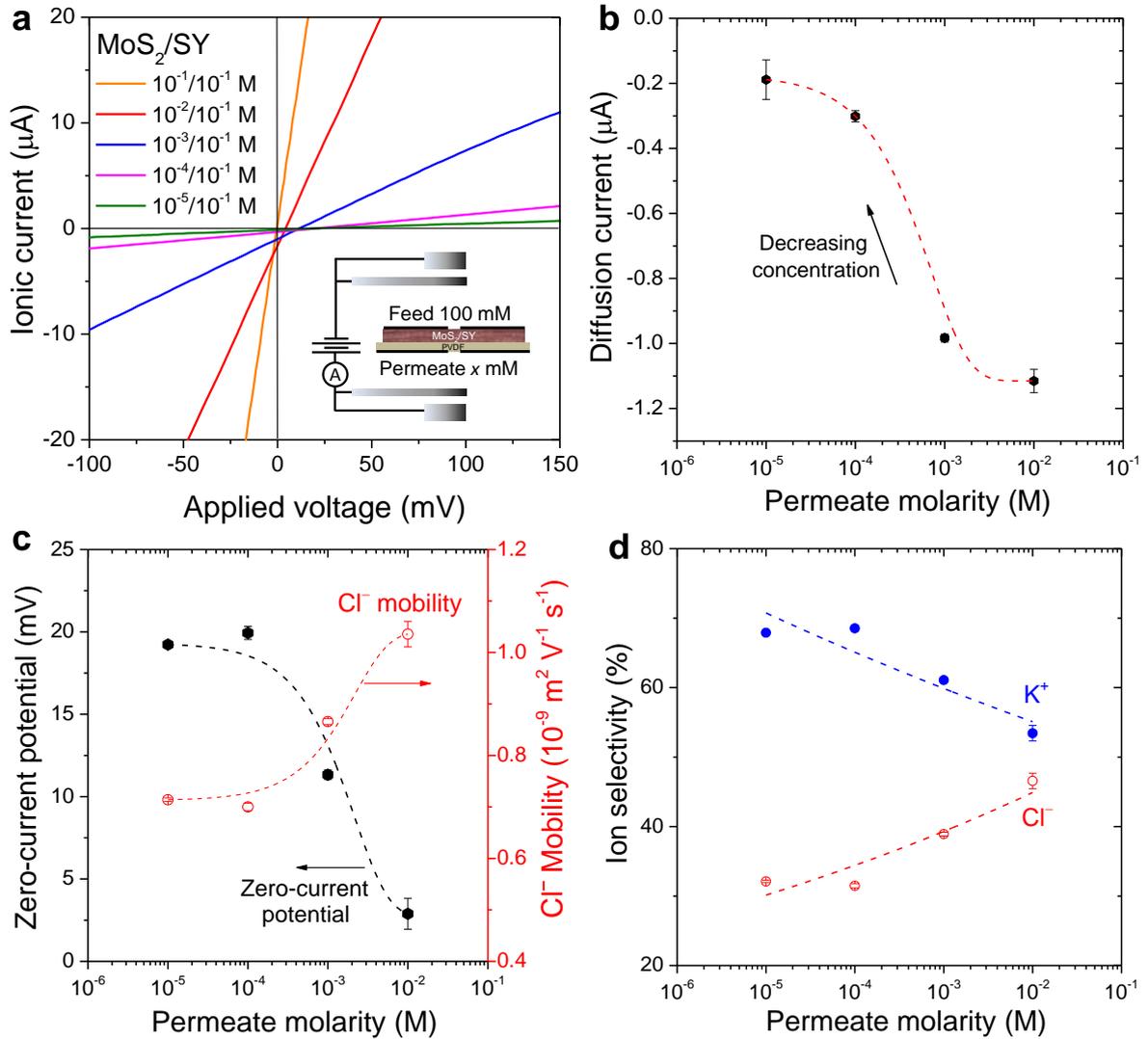

**Figure 3.** Molarity dependence of ion mobility. (a) *I–V* characteristics of the 3.01 ± 0.13 μm thick MoS$_2$/SY membrane for KCl solutions at several permeate concentrations using a constant KCl concentration at the feed reservoir (100 mM). The inset shows a schematic of the drift-diffusion experiment at different permeate molarities (*x* mM). (b) The net ion current as a function of KCl concentration at the permeate reservoir. (c) The zero-current potential (black line) and Cl$^-$ mobility (red line) as a function of the permeate molarity. The Cl$^-$ mobility was calculated using the combination of the GHK equation (Eq. 1) and the ionic conductivity (Eq. 2). (d) Ion selectivity (%*S*) of the MoS$_2$/SY membrane for K$^+$ and Cl$^-$ as a function of permeate molarity. Error bars indicate standard deviation in our measurements of the zero-current potentials and ionic conductivity.

To understand the role of the adsorption of hydronium and hydroxide ions on the nanocapillary channels, and the consequent effect this has on membrane stability, the *I-V* responses of the membranes were measured for a constant KCl concentration for both sides (100 mM) as various pH, from 3.4 to 10.8 as shown in Figure 4a. Figure 4b shows the ionic conductance of the dye functionalized $MoS_2$ membranes as a function of pH values. The KCl conductance of the functionalized membrane slightly decreased in acidic media which corresponds to low adsorption of $H^+$ on the surface of the $MoS_2$/SY channels. By contrast, the conductance increased sharply for basic media which indicated significantly higher $OH^-$ adsorption occurred on the $MoS_2$/SY surfaces, resulting in an increase in negative charge density inside the nanochannels (see Table S1 for bulk conductivity at various pH).[18, 38] Similar behavior with increasing conductance at high pH has been reported for GO membranes,[18] carbon nanotubes,[38] and hBN nanochannels.[25] Iso-pH conditions were maintained for both feed and permeate reservoirs for this drift-diffusion experiment. Figure 4c shows the $K^+/Cl^-$ mobility ratio of the functionalized $MoS_2$ membrane and a pristine $MoS_2$ membrane as a function of pH. A pristine exfoliated $MoS_2$ membrane exhibited a small change in mobility ratios for a range of pH solutions with slightly increased at basic pH ($K^+/Cl^-$ > 1) which indicated to partial adsorption of hydroxide ions on the surface of the pristine $MoS_2$ channels. Interestingly, the mobility ratio of the dye functionalized $MoS_2$ membrane significantly decreased at low pH values, which is consistent with some protonated $MoS_2$/SY surface corresponding to the decreasing negative surface charge of $MoS_2$/SY, as indicated by the zeta potential measurements (see Figure S9). By increasing pH values, the mobility ratio dramatically increased due to the greater extent of $OH^-$ adsorption at the functionalized sites of the $MoS_2$/SY surface. This significant change in mobility ratio indicated the altered surface chemistry of $MoS_2$ after functionalization that leads to the effectiveness of the membrane for nanofiltration with a high cation selectivity of ~80% for basic pH. Figure 4d shows the $K^+$ mobility increased and $Cl^-$ mobility decreased ($K^+/Cl^- \gg 1$) at high pH due to the influence of polarized water molecules around their hydration shells. This is attributed to preferential co-ordination of $H^+$ and $OH^-$ groups on the $K^+$ and $Cl^-$, respectively.[39] This corresponds to an increase in the $OH^-$ adsorption on the nanochannel's surface at higher pH (see Figure 4b); leading to a strong influence on the $Cl^-$ transport on the $MoS_2$/SY surfaces (higher $Cl^-$ friction).

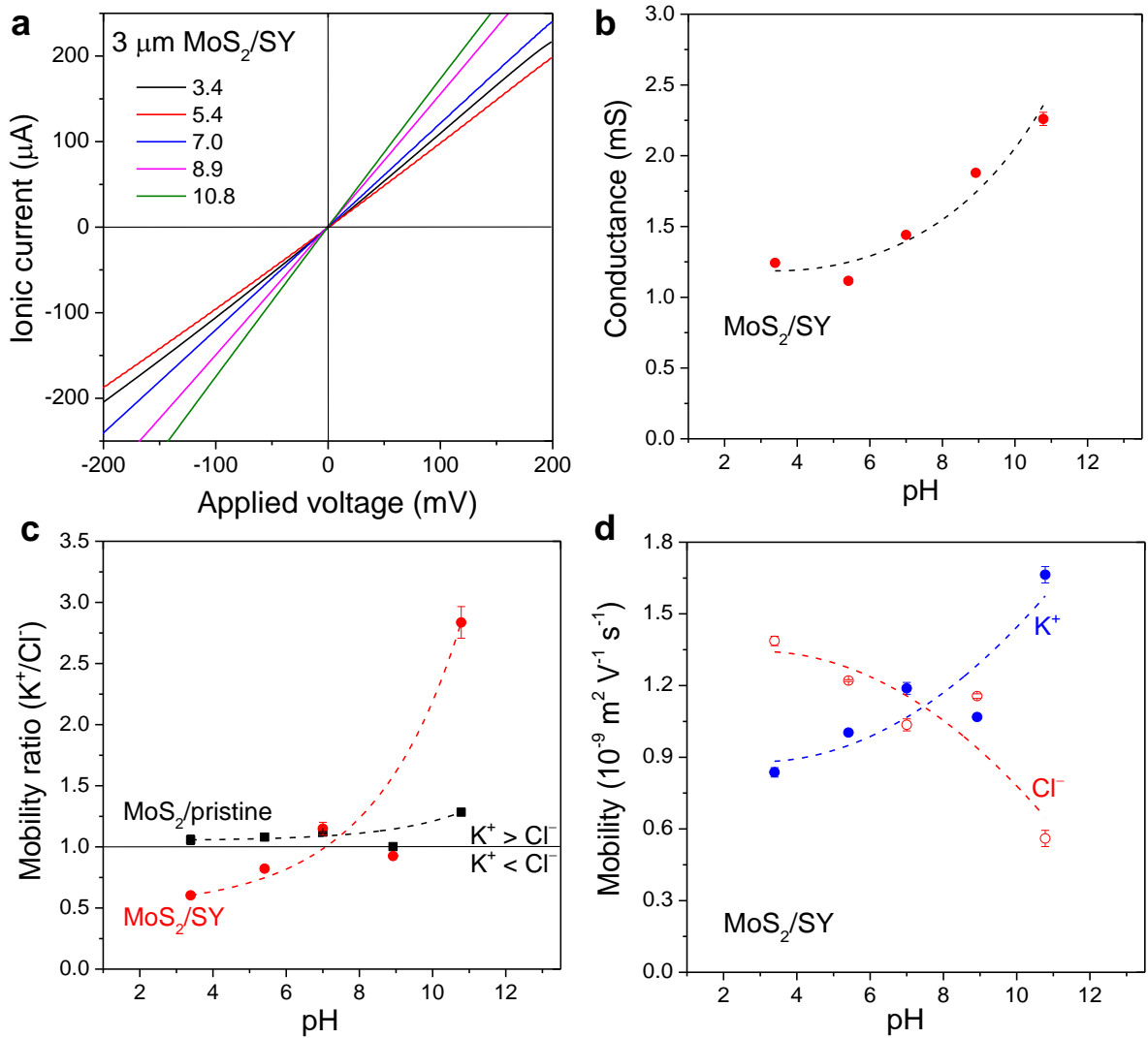

**Figure 4.** pH-dependent behaviour. (a) *I–V* characteristics of the 3.01 ± 0.13 μm thick $MoS_2$/SY measured at a constant KCl concentration of 100 mM for a variety of pH values,

from 3.4 to 10.8. (b) Corresponding conductance of the MoS$_2$/SY membranes as a function of pH values. (c) Mobility ratio, K$^+$/Cl$^-$, of the MoS$_2$/SY membranes and the pristine exfoliated MoS$_2$ membranes for several pH values. The black solid line indicates the equivalent ion mobility between K$^+$ and Cl$^-$. (d) K$^+$ and Cl$^-$ mobility of the MoS$_2$/SY membranes for different pH values. Note the additional ions of HCl and KOH solutions to achieve the required pH gave the small increase (< 5%) in the bulk electrolyte conductivity (see Table S1).

In addition, the change in surface chemistry after functionalization is reflected in changes in the water contact angle (WCA) in air, by using ultra-pure deionized water droplets placed on the surface of the MoS$_2$ membranes (see Figure S10). The WCA of the MoS$_2$ membrane after functionalization (~68°) was much lower than a fresh, pristine MoS$_2$ (~85°), i.e., the dye functionalized MoS$_2$ membrane is more hydrophilic. This is attributed to the dye functionalization altering the surface chemistry and roughness of the MoS$_2$. This change may result from the increase in dye functionalization on the Mo-edge of MoS$_2$ (see HAADF-STEM in Figure S5) due to the high number of defects created by the bath sonication process, which is evident by the presence of Mo−N bonding determined by Raman and X-ray photoelectron spectroscopy (XPS) as shown in Figure S11-12, respectively. Functionalized MoS$_2$ membranes were also soaked in different pH solutions (10$^{-3}$-10$^{-5}$ M for HCl and KOH solutions) for a week and dried in vacuum oven at room temperature. The functionalized MoS$_2$ membrane exhibits a constant WCA for the acid pH but slightly drops at the basic pH 10.8, which is again indicative of a more OH$^-$ adsorption giving more hydrophilic material.[40,41]

In conclusion, the dye functionalized MoS$_2$ membranes (MoS$_2$/SY) studied by the drift-diffusion technique show a strongly reduced ionic mobility, which is seen for a range of pH values due to the stability of the membrane materials. The ion mobility of the MoS$_2$/SY membrane was reduced by ca. 10-fold, compared to the pristine $MoS_2$ membrane and other commercial membranes. Moreover, the molar conductivity decreased by over a factor of ~5 and ~50 for $MoS_2$/pristine and $MoS_2$/SY, respectively, compared to the bulk molar conductivity. The treatment of dye functionalization of $MoS_2$ membranes not only achieved excellent charge/size selective ionic sieving but also enabled the tunability for cation selectivity of nearly 80% under basic pH. The $MoS_2$/SY membranes demonstrated significantly improved ion rejection compared to those previously reported for GO, $Ti_3C_2T_x$, and porous polymeric membranes. Thus, the ability to alter the surface charge of the membranes via functionalization with charged organic molecules provides the capability to modulate ion transport selectively for technologies used in water purifications such as ion exchange membranes, specifically in applications for electro-dialysis and electro-deionization.

**Supporting Information**

Materials and methods including silver/silver chloride agarose salt bridge preparation, membrane preparation and dye functionalization method; further membrane characterizations including SEM/EDX, STEM, XRD, Raman, XPS, zeta potential measurements, and water contact angle (WCA) experiments; further ion transport results through a pristine $MoS_2$ membrane and GO-based membranes including electrolyte and ionic conductivity measurements; ion mobility comparison between hydrated and non-hydrated cations; voltage drop across a bare PVDF membrane.

**Corresponding Author**

*Email: Robert.dryfe@manchester.ac.uk. *Email: Mark.bissett@manchester.ac.uk.

ACKNOWLEDGEMENTS

W.H. wished to acknowledge the Development and Promotion of Science and Technology Talents Project (DPST), Royal Government of Thailand scholarship. We also thank funding from Engineering and Physical Sciences Research Council, UK (grant references EP/R023034/1, EP/N032888/1 and EP/P00119X/1). SuperSTEM is the UK Engineering and Physical Sciences Research Council (EPSRC) National Research Facility for Advanced Electron Microscopy. All research data supporting this publication are directly available within this publication and the corresponding Supporting Information as well as available from the corresponding authors upon reasonable request.

# Supporting Information

# Potential Dependant Ionic Sieving through Functionalized Laminar MoS$_2$ Membranes

*Wisit Hirunpinyopas,[1,2] Eric Prestat,[3,4] Pawin Iamprasertkun,[1,2] Mark A. Bissett,[1,3]\* Robert A. W. Dryfe[1,2]\**

[1]National Graphene Institute, [2]School of Chemistry, and [3]School of Materials, University of Manchester, Oxford Road, Manchester, M13 9PL, United Kingdom

[4]SuperSTEM Laboratory, SciTech Daresbury Campus, Daresbury, WA4 4AD, United Kingdom

mark.bissett@manchester.ac.uk or robert.dryfe@manchester.ac.uk*

**Contents:**

1. Materials and methods
2. Membrane preparation and dye functionalization
3. Electron microscopy characterizations
4. X-ray diffraction analysis
5. Ion transport through a pristine MoS$_2$ membrane
6. Electrolyte and ionic conductivity measurements
7. Ion transport through GO-based membranes
8. Zeta potential measurements
9. Water contact angle (WCA) measurements
10. Raman analysis of MoS$_2$ membranes
11. XPS analysis of MoS$_2$ membranes
12. Comparison between inorganic and organic cations
13. Voltage drop across a bare PVDF membrane
14. Supporting References

## 1. Materials and methods

### 1.1. Materials

Agarose powder (bio-reagent grade), $MoS_2$ powder (99%, ~6 μm, max. 40 μm), and sunset yellow powder (disodium 6-hydroxy-5-[(4-sulfophenyl)azo]-2-naphthalenesulfonate; dye content 90%) were purchased from Sigma-Aldrich. Omnipore membrane filters (polyvinylidene fluoride (PVDF), hydrophilic, 0.1 μm pore size, and 13 mm diameter) were purchased from Merck Millipore Limited. All aqueous solutions were prepared from ultra-pure deionized water (Milli Q water purification system, 18.2 MΩ cm resistivity at room temperature).

### 1.2. Agarose salt bridge preparation

Firstly, silver wire (99.99 % purity, 0.35 mm diameter, Goodfellow Cambridge Limited: AG005145/1) was used to form a spring electrode to obtain a high surface area for chloride deposition. A silver wire spring was cleaned by immersing in 0.1 M $HNO_3$ solutions for a few seconds to remove any contamination from the surface and then rinsed with deionized water. Secondly, silver/silver chloride (Ag/AgCl) electrodes were prepared in 0.1 M HCl solution by applying the potential of 0.5, 1.5, and 2.5 V for an hour with a Pt flag used as the counter electrode. Then, the prepared Ag/AgCl electrodes were cleaned by rinsing with deionized water to remove any residue.[1] Finally, the stability of the as-prepared Ag/AgCl reference electrodes were tested as working and reference electrodes in 0.1 M KCl solution by applying current of 0 A for 1 hour. The stable Ag/AgCl reference electrodes showed a potential difference around −0.8 mV (ideally should be less than ±1.0 mV).

To maintain a stable electrode potential during the experiment, Ag/AgCl reference electrodes were prepared with agarose salt bridges inside glass luggin capillaries. Briefly, agarose (3%) was dissolved in 3 M KCl solution by heating in a water bath at 75-80 °C under constant stirring. The tip of luggin capillary was immersed into the hot agarose gel which allowed the gel to fill the luggin tube by capillary force and then filled the rest by glass pipette. The Ag/AgCl electrode was immediately immersed in the gel leaving 3-5 mm between the luggin tip and the end of the electrode. The agarose salt bridge was left for an hour to allow the gel to set completely.[2] Then, the top of the luggin tubes was sealed with epoxy glue. The resultant Ag/AgCl agarose salt bridges were kept in 3M KCl solution until use to avoid gel getting dried. The Ag/AgCl salt bridges were used in a custom-made H-beaker cell for the drift-diffusion experiment to eliminate the potential occurring from the redox reactions on the electrodes under different salt concentration gradients.

### 1.3. Cross-sectional STEM characterization methods

The cross-sectional scanning transmission electron microscope (STEM) images were carried out using an FEI Titan 80-200 ChemiSTEM equipped with probe-side aberration correction and an X-FEG electron source, which was used for the aberration-corrected scanning transmission electron microscope. STEM images were recorded using an acceleration voltage of 200 kV, a convergence angle of 21 mrad, and beam currents of 90 pA for imaging.

## 2. Membrane preparation and dye functionalization

### 2.1. Membrane preparation

MoS$_2$ dispersions were prepared by liquid phase exfoliation (LPE) using an ultra-sonication process.[3, 4] MoS$_2$ powder (1 g) in 100 mL of a mixture of isopropanol and water (1:1 v/v)[5] was sonicated with a frequency of 37 KHz and a power of 328 W for 12 hours at 15 °C as shown in Figure S1a. The dispersions were centrifuged twice at 6000 rpm (3139 g) for 30 minutes to remove any non-exfoliated material. The supernatant containing a distribution of flake sizes was separated from the sediment by removing the top 80% of the solution. Figure S1b shows the stable MoS$_2$ dispersion, which maintained high stability over 6 months. The MoS$_2$-based membranes were fabricated using a programmable syringe pump with an applied flow rate of 10 mL h$^{-1}$, using the PVDF membrane as support. The laminar MoS$_2$ membranes were prepared in different thicknesses (1, 3, and 5 µm) using the relation between the thickness and MoS$_2$ mass loading as described in our previous work.[6]

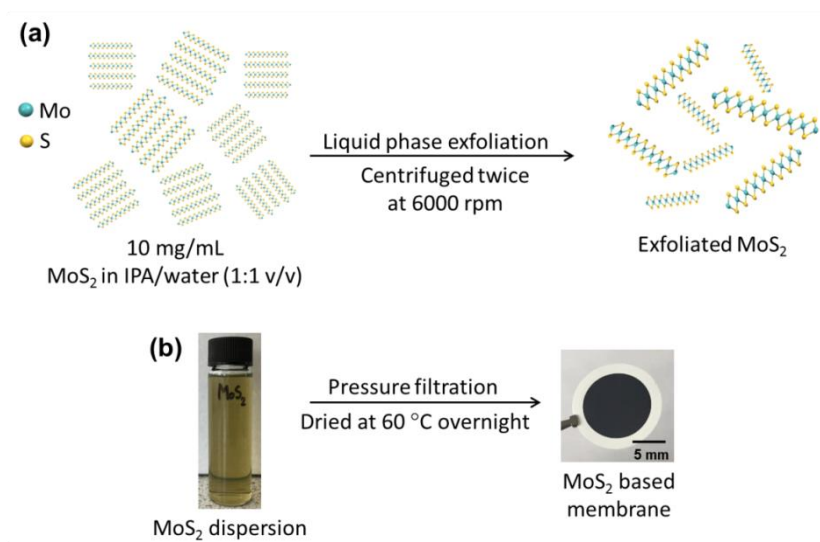

**Figure S1.** Exfoliation process and membrane preparation. (a) Exfoliation of MoS$_2$ in a mixed solvent of isopropanol and water (1:1 v/v) via bath sonication process. (b) Photographs of MoS$_2$ dispersion and membrane prepared by pressure filtration.

**2.2. Dye functionalization**

To control the exposed membrane area during the functionalization process, the $MoS_2$ membrane was assembled between polyethylene terephthalate (PET) sheets with epoxy resin to give an exposed area of 0.26 cm$^2$. Importantly, the $MoS_2$ sandwich membranes were first checked by optical microscopy to ensure there was no obvious physical damage (e.g. cracking, defects, and delamination) as shown in Figure 2a. The dye functionalization process was carried out as reported in our previous work.[6] Figures 2b-c shows the dye functionalization using 1 mM SY and ultra-pure water in the feed and the permeate sides. The maximum amount of SY sorbed onto the $MoS_2$ for complete monolayer coverage, calculated from Langmuir adsorption isotherm, was ~70 mg g$^{-1}$ of $MoS_2$.[6]

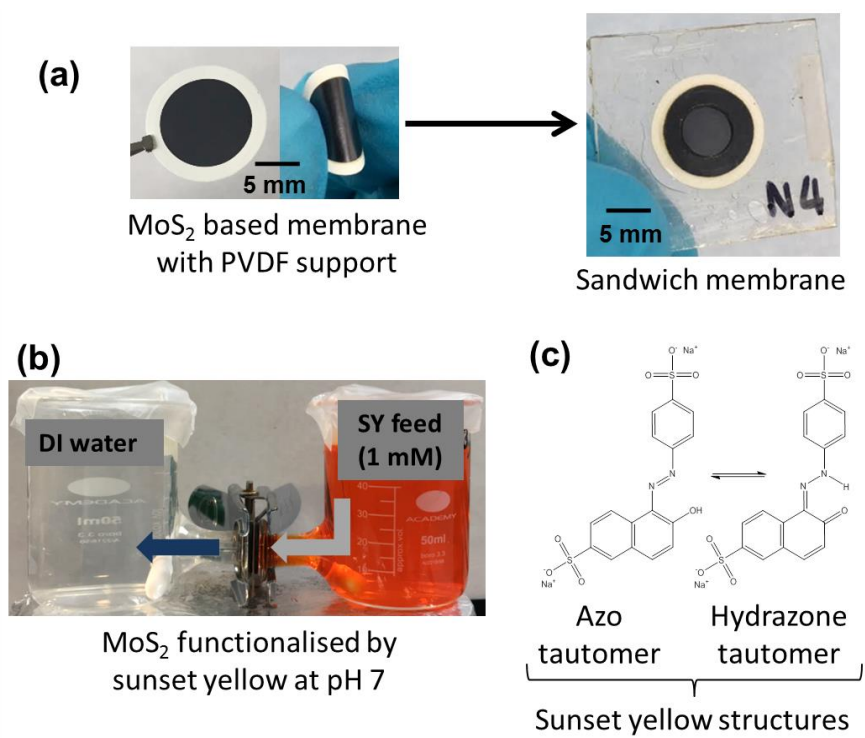

**Figure S2.** (a) Photographs of $MoS_2$ membrane with PVDF support with the resulting sandwich membrane. (b) Photographs showing dye functionalized $MoS_2$ membrane at neutral pH achieving the functionalization process from our previous work.[6] (c) The two possible molecular structures of sunset yellow showing azo and hydrazone tautomers.

## 3. Electron microscopy characterizations

### 3.1. SEM and EDX characterizations

Top-down and cross-sectional SEM images of the MoS$_2$/SY membranes, obtained using a FEI Quanta 650 FEG ESEM are shown in Figure S3-4. All SEM images were obtained with an accelerating voltage of 15 kV, under high vacuum conditions utilizing secondary electron detection. Figure S3 shows lateral MoS$_2$ flakes size in the range of 200-300 nm, with randomly stacked orientation. Further analysis of the cross-sectional SEM images was performed to gain insight in the morphology and the stacking of the MoS$_2$ flake as shown in Figure S4a-b. Figure S4c-d also show the Mo and S elemental maps obtained using energy dispersive X-ray (EDX) analysis. This confirms that the MoS$_2$ flakes are homogeneous across the MoS$_2$ membrane.

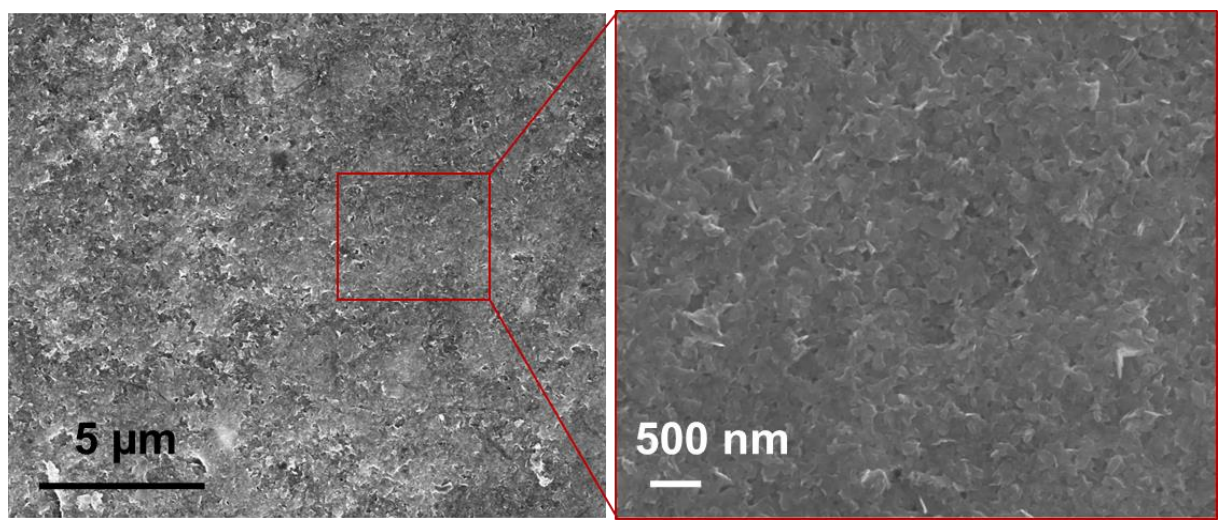

**Figure S3.** Top-down SEM images showing morphologies of the MoS$_2$/SY membranes.

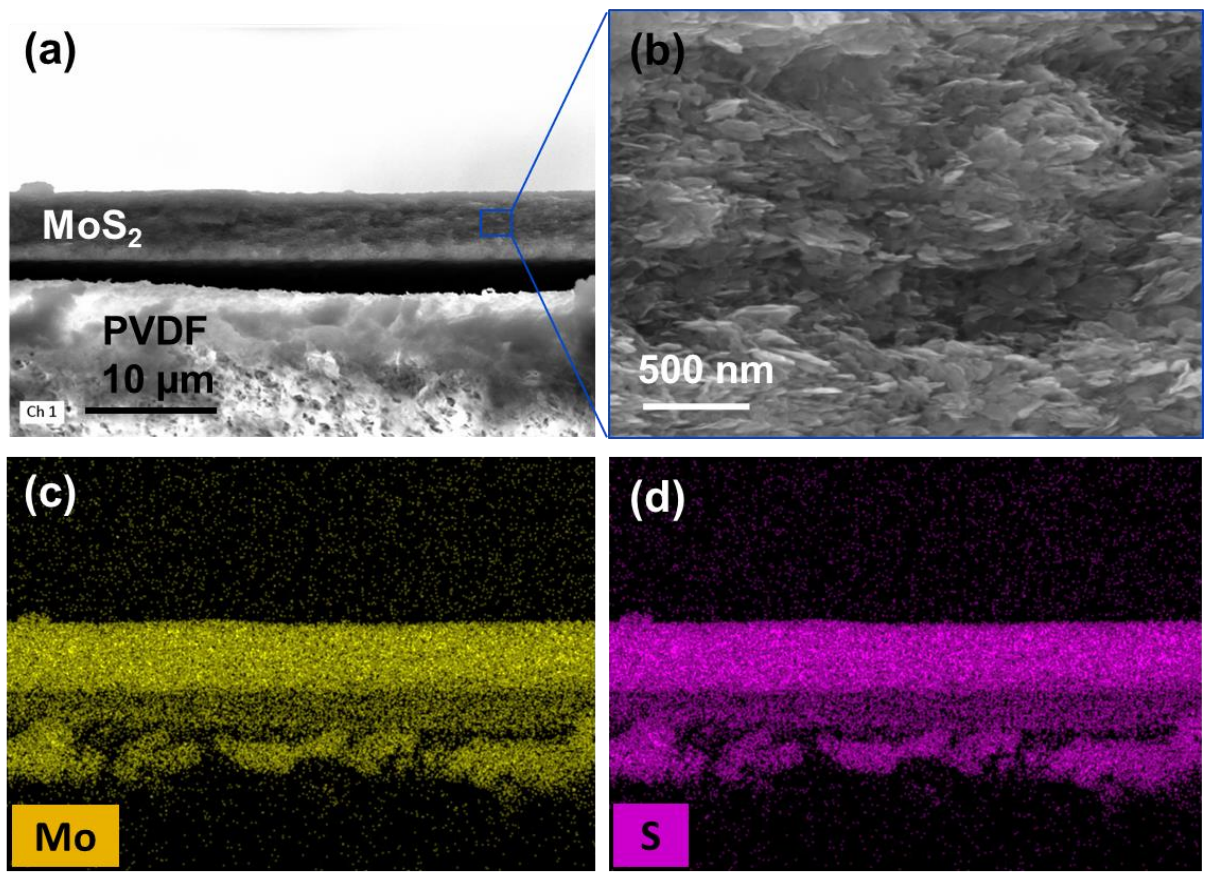

**Figure S4.** (a-b) Cross-sectional SEM images with element mapping for (c) Mo, and (d) S elements.

### 3.2. STEM characterization

High angle annular dark field scanning transmission electron microscope (HAADF-STEM) imaging was performed in near ultra-high vacuum (<1 × 10$^{-9}$ Torr) using an aberration corrected Nion UltraSTEM 100 operating at 60 kV and fitted with a cold field emission gun. The probe convergence angle was 30 mrad with a beam current of ~20 pA. The collection angle range of the HAADF detector was 86-190 mrad.[7]

Figure S5a shows the HAADF-STEM image of pristine exfoliated MoS$_2$ nanosheet. The thicknesses of the nanosheets are mostly 5 layers which is in agreement with a cross-section of the STEM image in the main text. High magnification of a monolayer MoS$_2$ area is also shown in Figure S5b; illustrating individual Mo and S atoms with trigonal prismatic coordination (see Figure S5c). The contrast of HAADF-STEM images is dependent on the atomic number and thus Mo atoms appear brighter than S atoms. Each Mo center of the 2H-phase MoS$_2$ is prismatically coordinated to six surrounding S atoms in the upper layer. The 2H-phase has a stacking order of AbA BaB AbA ..., which confirms the structure of the 2H-phase.[7, 8]

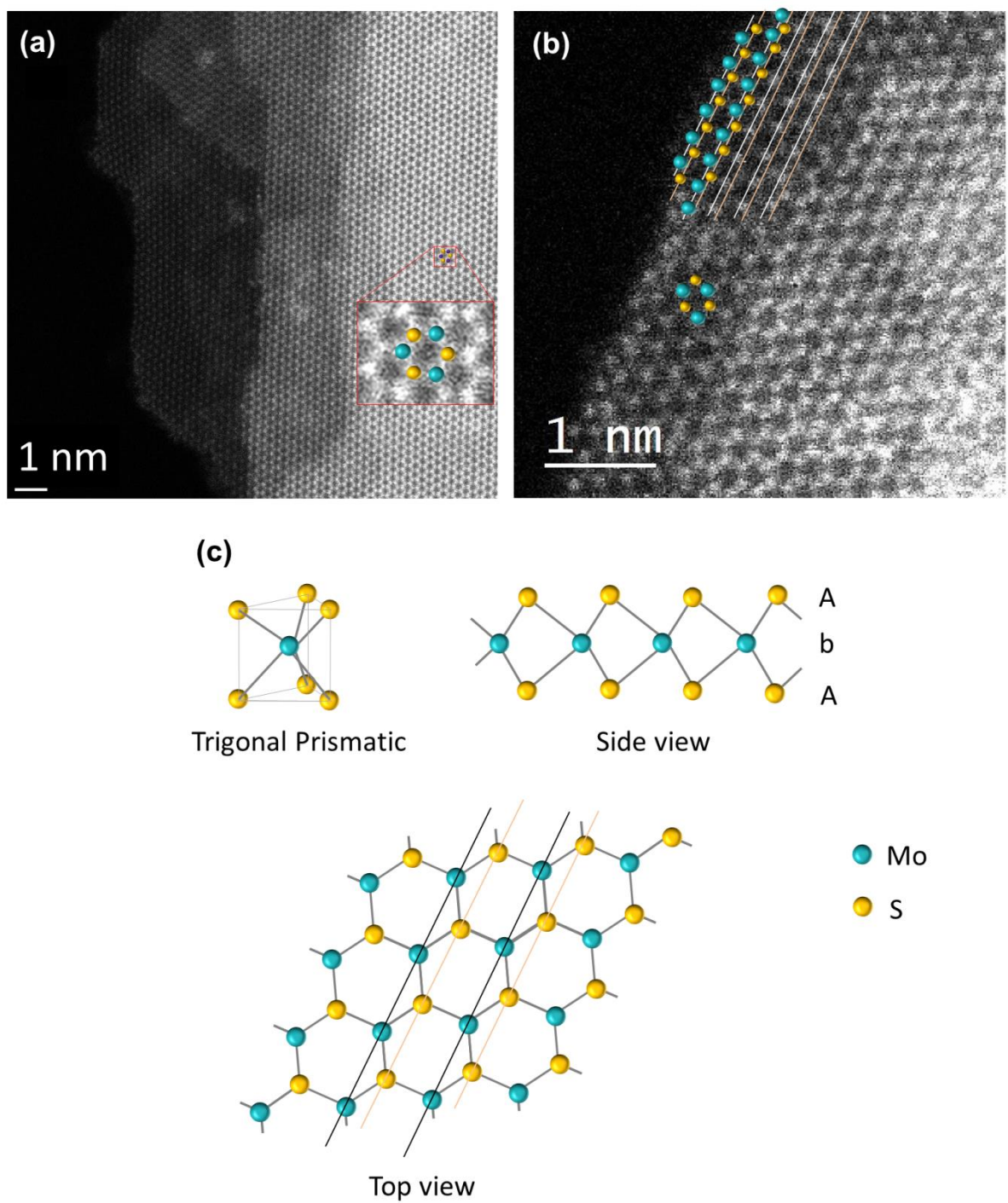

**Figure S5.** (a) HAADF-STEM image of a pristine exfoliated MoS$_2$ nanosheet showing the thickness consisting of ~5 layers. (b) HAADF-STEM image of monolayer pristine MoS$_2$ nanosheet illustrating Mo and S atoms. (c) Side and top views of the 2H structure for the MoS$_2$ layer. The trigonal prismatic coordination for the Mo atom in 2H-phase MoS$_2$ is also shown. The Mo and S atoms are green and yellow spheres, respectively.

**4. X-ray diffraction analysis**

Powder X-ray Diffraction (PXRD) patterns of the $MoS_2$ membranes were obtained using a PANalytical X'pert X-ray diffractometer. The patterns were recorded in the range $2\theta$ = 5-70°, with a step size of 0.017° with a scan step time of 66 s, which used a Cu-K$\alpha$ radiation source (0.154 nm wavelength) operating at 40 kV and 30 mA. The (002) peak positions of $MoS_2$ were corrected using the PVDF peak at $2\theta$ of 20.17° as an internal reference peak.

Figure S6 shows PXRD patterns of pristine and dye functionalized $MoS_2$ ($MoS_2$/SY) at the (002) peak supported on PVDF membranes. It is clear that the diffraction peak due to the (002) peak of $MoS_2$ after exfoliation is notably broader, compared to the bulk material, indicating the distribution in the number of layers and flake size from the ultrasonication process.[9, 10] The PXRD pattern of $MoS_2$/SY exhibited small change in the (002) peak position indicating that there was no significant swelling as the membranes have been exposed in aqueous solution for several months.[6, 11]

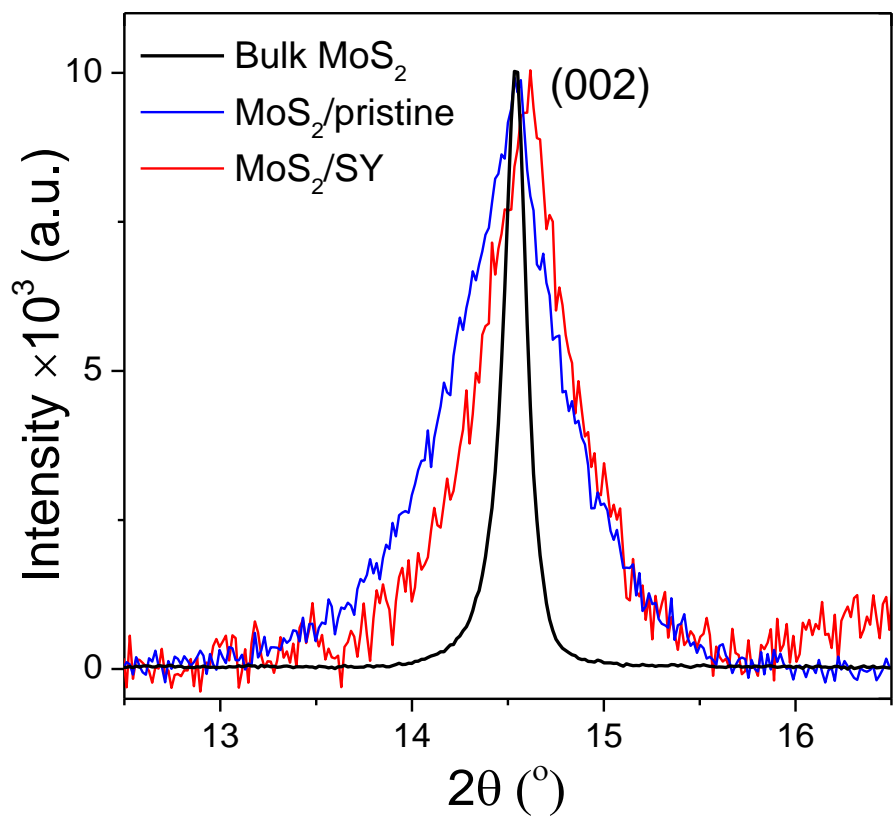

**Figure S6.** PXRD pattern of the (002) peak position for the $MoS_2$/SY, pristine exfoliated $MoS_2$, and the bulk materials.

## 5. Ion transport through a pristine MoS₂ membrane

Figure S7 shows the *I-V* characteristics of different valence ions (KCl, BaCl$_2$, and AlCl$_3$) of pristine exfoliated MoS$_2$ membrane (~3 μm thick), which were measured at a constant concentration ratio (10 mM/100 mM). It was shown that the zero-current potential significantly decreased to a more negative potential with increasing hydrated cation radii from K$^+$ to Ba$^{2+}$, but slightly decreased for Al$^{3+}$. This is because less charge and size selective ion sieving for trivalent cations through MoS$_2$/pristine when compared to MoS$_2$/SY as shown in Figure 2 in the main text.

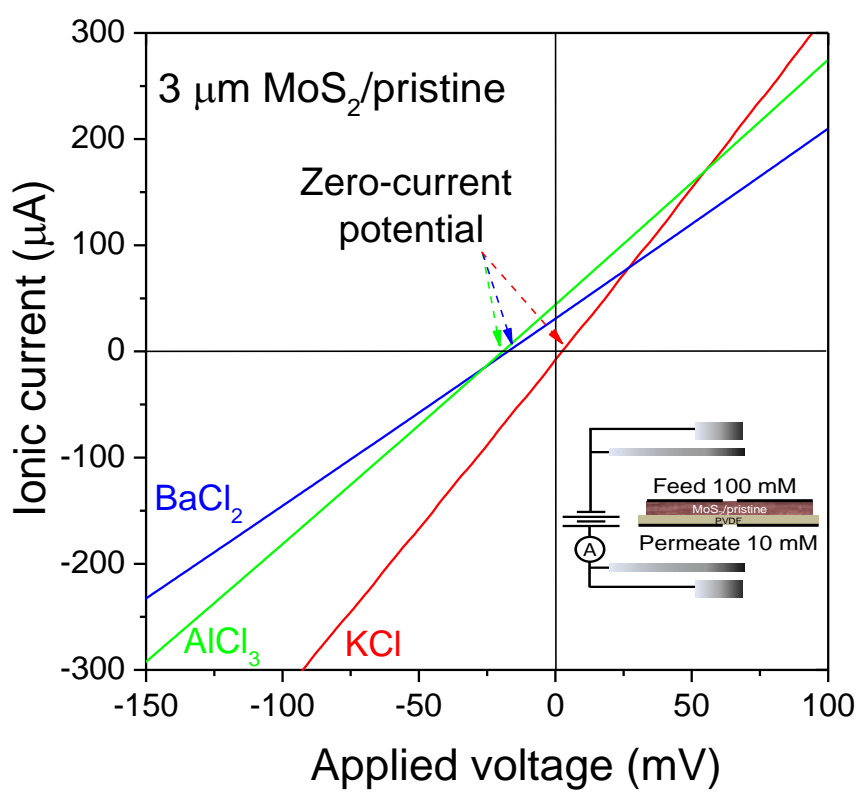

**Figure S7.** *I-V* characteristics of a pristine exfoliated MoS$_2$ membrane (3 μm thick), showing three different ionic salt electrolytes (KCl, BaCl$_2$, and AlCl$_3$) under the constant ratio between the feed (100 mM) and the permeate sides (10 mM). The inset shows a schematic of the drift-diffusion experiment for a pristine exfoliated MoS$_2$ membrane.

## 6. Electrolyte and ionic conductivity measurements

### 6.1 Electrolyte conductivity measurements

**Table S1.** Comparison of the measured electrolyte conductivity values in aqueous solution, reported at 25 °C.

| Electrolyte (0.1 M) | Electrolyte conductivity (mS cm$^{-1}$) | | | |
| --- | --- | --- | --- | --- |
| | Bulk conductivity | | MoS$_2$ membranes (~3 μm thick)[b] | |
| | Literature[12-14] | Measured[a] | MoS$_2$/pristine | MoS$_2$/SY |
| KCl | 12.89 | 13.18 | 2.43 | 0.215 |
| NaCl | 10.67 | 10.81 | 1.77 | 0.189 |
| LiCl | 9.58 | 9.78 | 1.67 | 0.169 |
| BaCl$_2$ | - | 20.10 | 3.29 | 0.352 |
| 1/2BaCl$_2$ | 10.51 | 10.68 | - | - |
| CaCl$_2$ | - | 19.12 | 3.36 | 0.345 |
| 1/2CaCl$_2$ | 10.24 | 10.28 | - | - |
| MgCl$_2$ | - | 18.52 | 3.35 | 0.340 |
| 1/2MgCl$_2$ | 9.71 | 10.00 | - | - |
| CeCl$_3$ | - | 26.67 | 4.91 | 0.639 |
| CrCl$_3$ | - | 26.27 | 4.95 | 0.612 |
| AlCl$_3$ | - | 25.30 | 5.21 | 0.448 |
| pH-dependent behavior | | | | |
| KCl pH 3.4 | | 13.83 | 2.47 | 0.282 |
| KCl pH 5.4 | | 13.51 | 2.42 | 0.254 |
| KCl pH 7.0 | | 13.18 | 2.43 | 0.327 |
| KCl pH 8.9 | | 13.38 | 2.33 | 0.427 |
| KCl pH 10.8 | | 13.61 | 2.73 | 0.514 |

[a] The ionic conductivity of bulk chloride solutions were measured by a conductivity meter.

[b] The ionic conductivity of the MoS$_2$ membranes were experimentally calculated by the conductance of each ion after subtracting the conductance measured for the same electrolyte in a bare PVDF membrane. On the basis of electrical conductance, it can be present as a reciprocal value to resistance. Then, the total conductance of a series circuit between a

laminar MoS$_2$ membrane and a bare PVDF ($G_{MoS_2|PVDF}$) can be calculated from the individual conductance contributing from a laminar stacked MoS$_2$ ($G_{MoS_2}$) and a bare PVDF ($G_{PVDF}$) following equation:

$$\frac{1}{G_{MoS_2|PVDF}} = \frac{1}{G_{MoS_2}} + \frac{1}{G_{PVDF}} \tag{S1}$$

By using this equation, we can calculate the conductance contributing from solely MoS$_2$ membrane as following:

$$G_{MoS_2} = \frac{G_{PVDF} G_{MoS_2|PVDF}}{G_{PVDF} - G_{MoS_2|PVDF}} \tag{S2}$$

To gain more information about individual ion mobility, the ion conductivity ($\sigma_{MoS_2}$) can be calculated from the measured conductance ($G_{MoS_2}$) as following:

$$G_{MoS_2} = \sigma_{MoS_2} \frac{A_{MoS_2}}{l_{total}} \tag{S3}$$

where $A_{MoS_2}$ is the exposed MoS$_2$ membrane area that perpendicular to the direction of the electrical current and $l_{total}$ is the total length of each nanochannel within the laminar MoS$_2$ membranes.

To estimate the total length of each effective nanochannel ($l_{total}$), we used the possible permeation path for ions through laminar stacked MoS$_2$ membranes as suggested by Nair and Geim.[15] Schematic S1 shows the laminar stacking structure of a MoS$_2$ membrane which was separated by the distance between individual MoS$_2$ sheets (nanochannel height, $d$). The nanochannel height was estimated in the range of ~6-13 Å (mostly 10 Å) measured by the cross-sectional STEM images, as shown in Figure 1e in the main text. This estimated channel height is slightly larger than the free spacing between interlayer $MoS_2$ distances (~6 Å) after deducting the thickness of a single $MoS_2$ sheet as reported by Wang and Mi.[16]

The possible path of ion transport through a $MoS_2$ membrane of thickness ($h$) can be estimated by a number of ion turns ($N$) as following:

$$N = \frac{h}{d} \qquad (S4)$$

where each turn ($N$) is equivalent to the length of the nanocapillary channel ($L$) which is measured around 200-300 nm by SEM and TEM images (see Figure S3-5). According to this estimation, each $MoS_2$ crystallites were assumed to be the same dimensions (length, width, and thickness). Therefore, the total length ($l_{total}$) of each effective nanochannel can be estimated by the following equation:

$$l_{total} = \frac{h}{d} L \qquad (S5)$$

For our typical 3 µm thick $MoS_2$ membranes, $l_{total}$ can be estimated approximately 0.6 mm for permeation path of ions through laminar $MoS_2$ membranes at which channel height and nanosheet length are 10 Å and 200 nm, respectively.

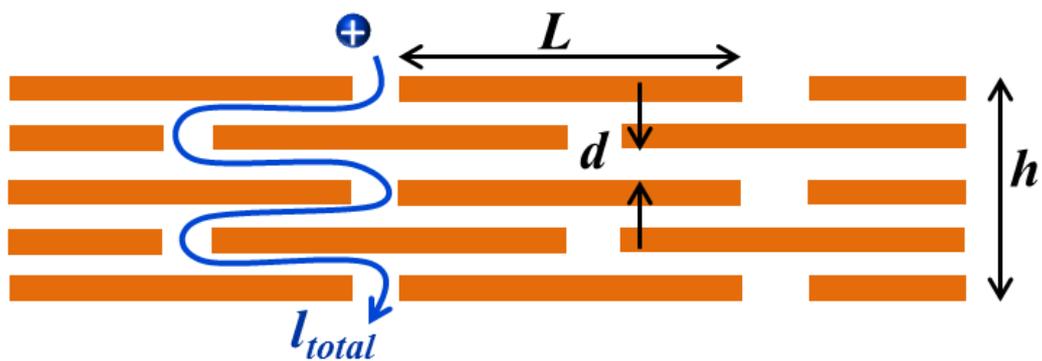

**Schematic S1.** Possible permeation path of ions inside the laminar stacked MoS$_2$ membranes.

**6.2 Ionic conductivity measurements**

**Table S2.** Literature values of molar (equivalent) conductivities and the calculated ionic conductivities of the MoS$_2$ membranes in aqueous solutions, reported at 25 °C.

| Ion | Molar conductivity (10$^{-4}$ m$^2$ S mol$^{-1}$) | | |
| --- | --- | --- | --- |
| | Bulk[13, 17] | MoS$_2$ membranes (~3 μm thick)[a] | |
| | | MoS$_2$/pristine | MoS$_2$/SY |
| K$^+$ | 73.5 | 12.8 | 1.15 |
| Cl$^-$ | 76.3 | 11.5[b] | 1.00[b] |
| Na$^+$ | 50.1 | 7.23 | 0.838 |
| Li$^+$ | 38.7 | 5.43 | 0.613 |
| Ba$^{2+}$ | 63.6 | 4.45 | 0.453 |
| Ca$^{2+}$ | 59.5 | 4.76 | 0.406 |
| Mg$^{2+}$ | 53.0 | 3.52 | 0.371 |
| Ce$^{3+}$ | 69.8 | 3.92 | 0.275 |
| Cr$^{3+}$ | 67.0 | 3.57 | 0.239 |
| Al$^{3+}$ | 61.0 | 3.54 | 0.144 |

[a] The molar ionic conductivity ($\lambda_i$) of the MoS$_2$ membranes were calculated using the following equation:[12, 17]

$$\lambda_i = F\mu_i \quad (S6)$$

where $\mu_i$ is the ion mobility of ion $i$ as reported in Figure 2b and $F$ is Faraday's constant.

[b] The molar Cl$^-$ conductivities were calculated using the Cl$^-$ mobility of KCl solutions.

## 7. Ion transport through GO-based membranes

To compare ion sieving performance with other laminar 2D material-based membranes, we also prepared GO membranes on a PVDF support with a comparable thickness to the $MoS_2$ membranes (~3 µm thick). The GO dispersion used herein was provided with the preparation method as reported in Wang and Dryfe.[18]

Figure S8a shows the *I-V* characteristics of KCl measured at a concentration gradient (10 mM/100 mM) of a GO membrane under the same testing conditions used to study the $MoS_2$ membranes. It was evident that the zero current potential significantly increased to a more positive potential of +31.3 mV which exhibited a high $K^+/Cl^-$ mobility ratio ca. 5, calculated by the GHK equation as explained in the main text. The mobility ratio from our prepared GO membrane was closed to the literature value reported by Hong et al.[19] with the percentage difference of <10 % as shown in Figure S8b.

Moreover, the individual mobility between $K^+$ and $Cl^-$ can be also estimated using the same method for the $MoS_2$ membranes. The permeation path for ions through the GO membrane was estimated to be ca. 1.5 mm at which the length of nanocapillary GO channel (*L*) and *d*-spacing (*d*) are 1 µm and 20 Å, respectively. The *d*-spacing used here was taken from the literature value when the membrane is immersed in water about an hour.[16] Therefore, the individual $K^+$ and $Cl^-$ mobility (●○ symbols) can be calculated as shown in Figure S8b.

To compare the estimated ion mobility of our prepared GO membrane with the literature value, the ion permeability ($P_i$) reported by Hong et al.[19] typically related to the diffusion coefficient ($D_i$) and the solubility of the ion transport through the membranes as defined by molar flux density (permeation rate) and Fick's first law, as following equation:

$$P_i = \frac{\beta_i D_i}{l_{total}}  \quad (S7)$$

According to the Nernst-Einstein relation, the ion mobility is related to the diffusion coefficient as expressed following[17, 20]

$$D_i = \frac{RT}{F}\mu_i  \quad (S8)$$

By combining Eq. S7 and S8, the ion mobility can be related to permeability as the following equation:

$$\mu_i = \frac{F}{RT}\frac{P_i l_{total}}{\beta_i}  \quad (S9)$$

where $\beta_i$ is the partition coefficient of an ion within the membrane. According to our results for ion mobility within the prepared GO membranes, $\beta_i$ can be assumed to be ca. 10 to evaluate $K^+$ and $Cl^-$ mobility (▲△ symbols) within the previously reported GO membrane.[19] In the case of the $Ti_3C_2T_x$ (MXene) membrane, as shown in Figure 2b, the $K^+$ mobility was also estimated from the permeation rate using equation S9 with the $l_{total}$ and $\beta_i$ were ca. 1.5 mm and 10, respectively. The $\beta_i$ of MXene membrane was assumed to be ca. 10 due to the equivalent ion permeability within $Ti_3C_2T_x$ and GO membranes as reported by Ren et al.[21]

To confirm the stability of GO membranes in aqueous media, Figure S8c shows a fresh dried GO membrane with a PVDF support prepared by pressure filtration at around 3 μm thick. It was clearly seen that the membrane exhibited significant instability from swelling after being immersed in deionized water for around 1 hour. This is in agreement with the previous literature using GO membranes for water purification.[16, 19, 22, 23]

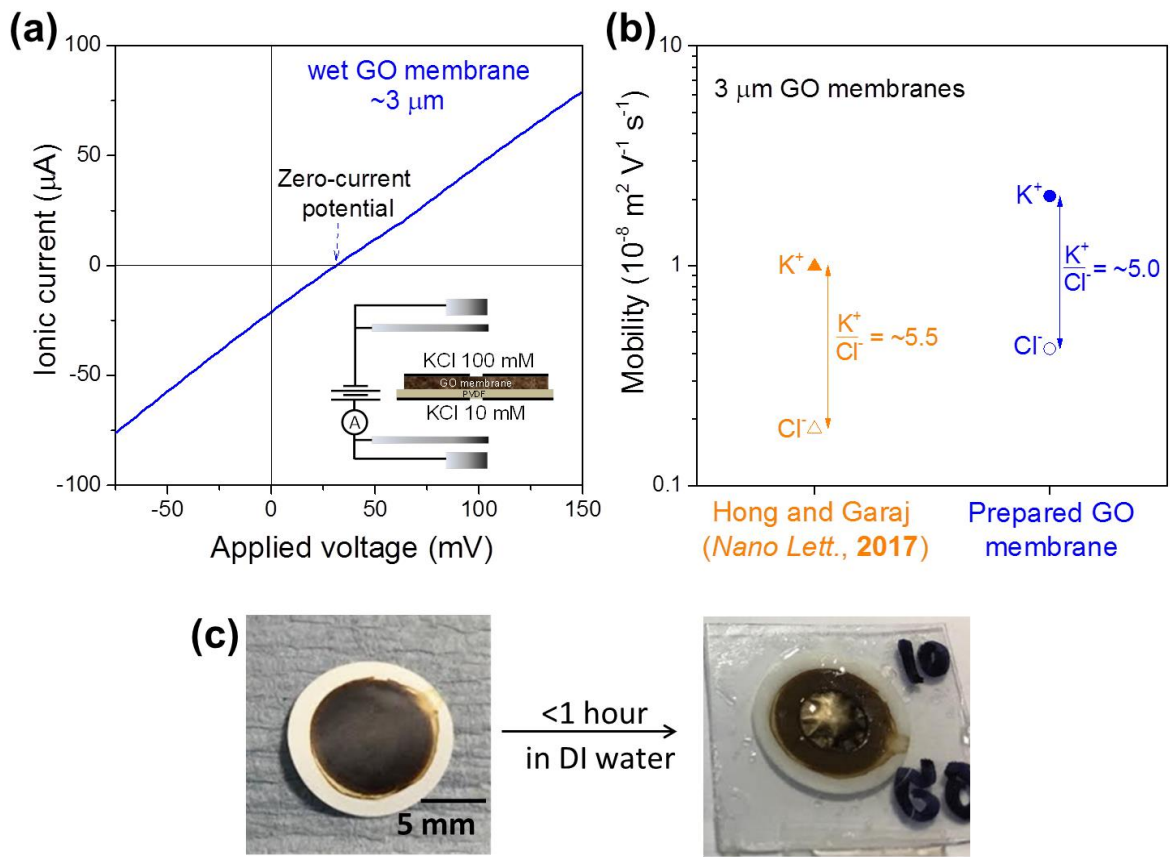

**Figure S8.** Ion transport through GO-based membranes. (a) *I-V* characteristic of KCl for as-prepared GO membrane with a comparable thickness of the MoS$_2$ membranes (~3 μm thick), measured under the concentration ratio (10 mM/100 mM). (b) Ion mobility of K$^+$ (●▲ symbols) and Cl$^-$ (○△ symbols) for our prepared GO membranes and the GO membrane as reported by Hong et al.[19] with the same thickness. (c) Photographs of a dried GO membrane at 3 μm thick and after immersed in deionized water for 1 hour.

## 8. Zeta potential of MoS$_2$ membranes

The surface charge of the MoS$_2$ was assessed by measurement of the zeta potential ($\zeta$) using a Malvern Nanosizer Z (NIBS) with Zeta-sizer software. The Smoluchowski equation was used for the determination of zeta potentials.[24]

The individual MoS$_2$ dispersions were prepared by re-dispersion of pristine MoS$_2$ membranes and dye functionalized MoS$_2$ membranes at various pH solutions ($10^{-3}$ and $10^{-5}$ M for HCl and KOH solutions) under the bath sonication for 15 minutes. These MoS$_2$ dispersions are well-dispersed for a range of pH solutions. Figure S9 shows the surface zeta potential of MoS$_2$ dispersion at different pH solutions. The resulting zeta potentials were measured at least three times. The resulting zeta potential at lower and higher pH values are in agreement with the previous work using a pristine MoS$_2$ and GO dispersions.[22, 25]

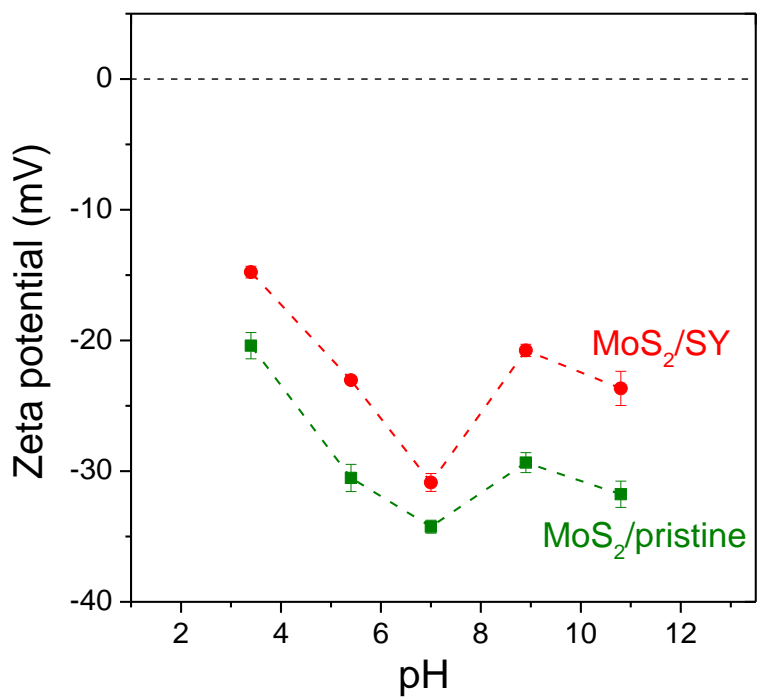

**Figure S9.** Zeta potential ($\zeta$) of the MoS$_2$/SY membrane and the pristine exfoliated MoS$_2$ membrane as a function of solution pH.

## 9. Water contact angle (WCA) measurements

A Theta Optical Tensiometer (Biolin Scientific, Finland) was used for measurement of water contact angle (WCA) in air with OneAttension software (version 2.3) in the sessile mode to control the water droplets and contact angle values.

The WCA of pristine $MoS_2$ and dye functionalized $MoS_2$ ($MoS_2$/SY) at different pH values are shown in Figure S10, recorded one second after placement of the water droplet. The water droplets for all membranes were controlled to a volume of ~6 μL, inside a high humidity vessel to control water droplet evaporation. The WCAs were recorded for 10 minutes using a camera to capture the droplet images at 1.9 FPS. The WCAs were calculated using the Young-Laplace equation.[26, 27]

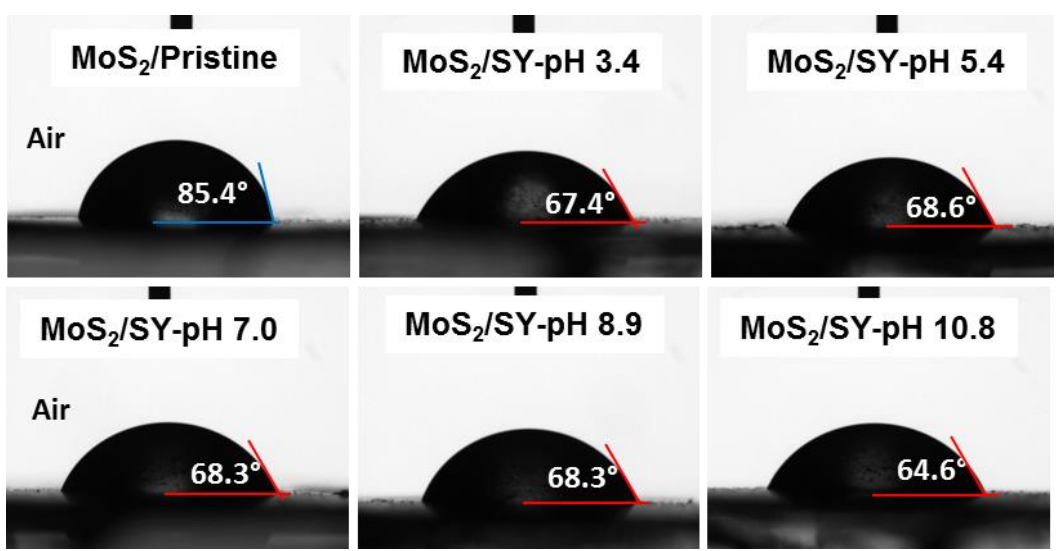

**Figure S10**. Photographs of the water contact angle (WCA) on a pristine exfoliated $MoS_2$ and dye functionalized $MoS_2$ ($MoS_2$/SY) membranes at different pH values, one second after placement of the water droplet. The experiment was repeated three times by changing the position of $MoS_2$ surfaces.

## 10. Raman analysis of MoS$_2$ membranes

Raman spectra were measured using a Renishaw inVia microscope with a 532 nm (2.33 eV) laser, incident perpendicular to the membranes. The laser power was 1 mW with a 50× objective lens, and a grating of 1800 l/mm. Spectra were obtained between 200-1700 cm$^{-1}$ and averaged over 3 accumulations.

Figure S11 compares the Raman spectra of the functionalized MoS$_2$ membrane with pristine exfoliated MoS$_2$, showing the characteristic E$_{2g}$ and A$_{1g}$ Raman modes of 2H-MoS$_2$ as labelled, and the vibrational modes for the raw dye (SY) powder with its two possible tautomeric structures (the N=N azo and N—H hydrazone).[28-30] The frequencies of characteristic MoS$_2$ peaks were shifted to lower energies around ~0.7 and ~0.6 cm$^{-1}$ (measured from ~20 Raman spectra) for E$_{2g}$ and A$_{1g}$, respectively, which indicates new bonding between MoS$_2$ and dye corresponding to XPS analysis (discussed later). Moreover, many vibrational bands which are characteristic of SY are not present after functionalization. The Raman spectra of the azo compound containing a N=N bond shows a strong band assigned to stretching mode around 1300–1580 cm$^{-1}$.[30] The strong bands observed at ~1320 – 1450 and 1596 cm$^{-1}$ for pure SY powder are characteristic of di-substituted naphthalene and phenyl C—C bond stretching, respectively, whereas those bands are completely absent after the functionalization process.[29, 30] This can be attributed to the chemical interaction between the dye and MoS$_2$ resulting in the disappearance of vibrational modes contributing from the SY structure. Moreover, the Raman spectrum of MoS$_2$/SY shows a new broad peak at 1152 cm$^{-1}$ attributed to bonding between Mo and N.[31]

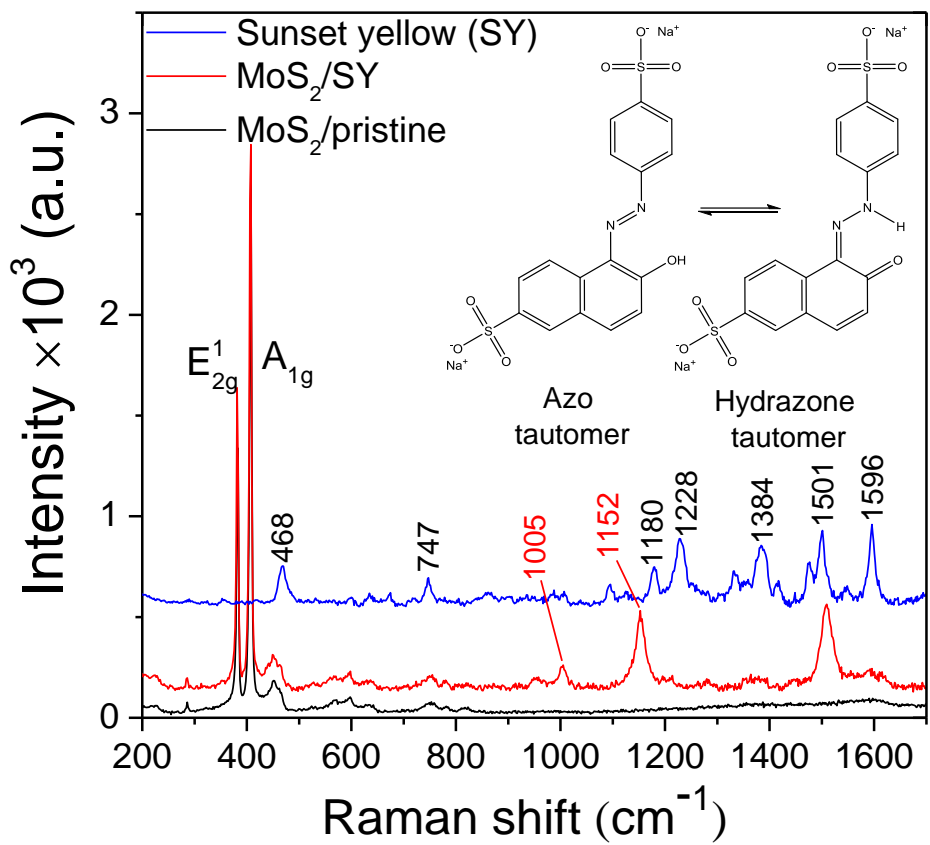

**Figure S11.** Raman spectra of a pristine exfoliated $MoS_2$ and the $MoS_2$/SY membranes compared to SY (raw powder). Inset shows two possible tautomeric SY structures.

## 11. XPS analysis of MoS$_2$ membranes

X-ray photoelectron spectroscopy (XPS) analysis was performed with a Kratos Axis Ultra spectrometer, with excitation from a focused monochromated Al Kα source (1486.6 eV) and using an electron flood gun for charge neutralization. All XPS spectra were calibrated using adventitious carbon (C 1s) at 284.8 eV. Peak fitting used a nonlinear Shirley-type background (70% Gaussian and 30% Lorentzian line shapes).

The XPS analysis is shown in Figure S12. The Mo 3d spectrum of the pristine exfoliated MoS$_2$ membrane consists of two main component peaks at around 230 and 233 eV assigned to Mo$^{4+}$ 3d$_{5/2}$ and Mo$^{4+}$ 3d$_{3/2}$ for 2H–phase MoS$_2$, respectively (Figure S12a). The Mo 3d peaks are shifted to lower binding energies by ~0.6 eV after functionalization which is similar to the S 2p regions of the spectra. This is due to charge transfer of new bound systems for Mo atoms. The peaks assigned at ~395 and ~413 eV correspond to Mo 3p$_{3/2}$ and Mo 3p$_{1/2}$, respectively. The N 1s peak at ~400 eV shown in Figure S12b for MoS$_2$/SY can also be confirmed as strongly chemisorbed nitrogen from azo/hydrazone species on the metallic molybdenum atoms.[32-34] This was also supported by Becue et al.[33] who reported a value of ~400–402 eV for Mo─N bonding.

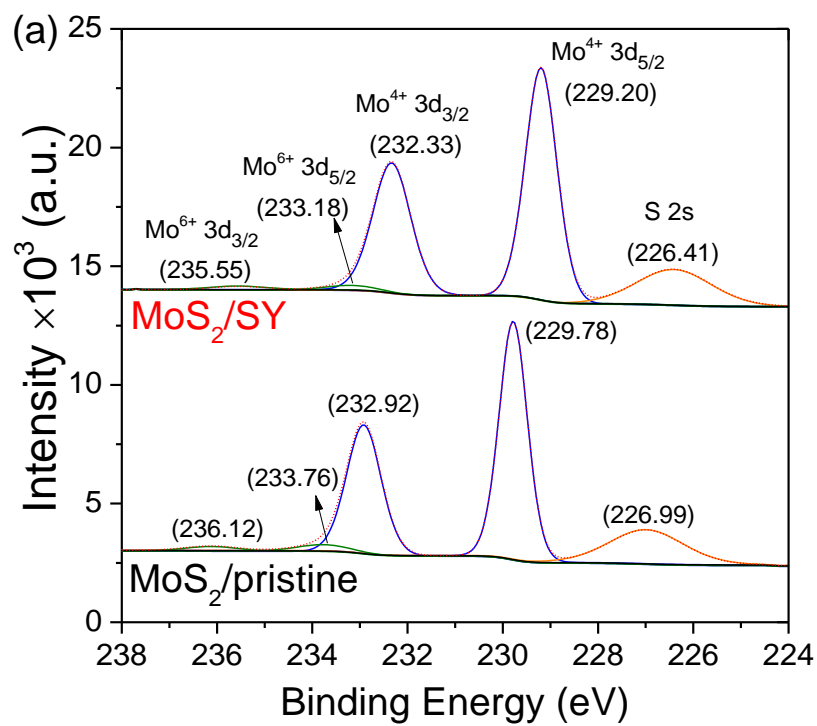

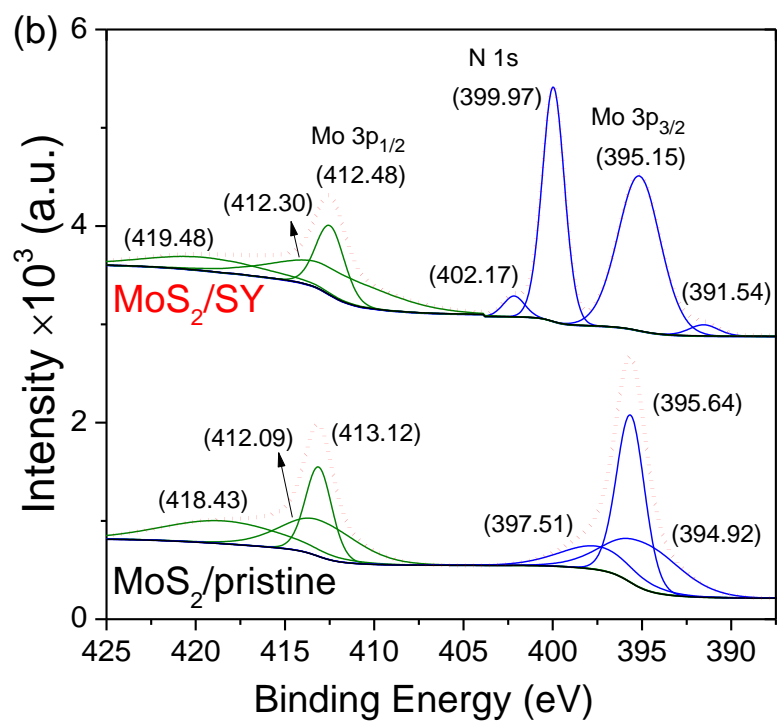

**Figure S12.** High-resolution XPS spectra of the Mo 3d (a) and Mo 3p (b) regions for pristine exfoliated $MoS_2$ and functionalized $MoS_2$ ($MoS_2$/SY) membranes.

**12. Comparison between inorganic and organic cations**

Furthermore, mobility measurements with the functionalized $MoS_2$ membrane ($MoS_2$/SY) were also carried out with organic cations, specifically tetramethyl-, tetraethyl-, and tetrapropylammonium ions (present as their chloride salts TMA-Cl, TEA-Cl, and TPrA-Cl) to make a comparison with the inorganic cations (NaCl, $BaCl_2$, and $CeCl_3$), which have hydrated radii very close to the radii of the non-hydrated organic ions.[35-37] The *I-V* characteristics for both types of salt were measured with a constant concentration ratio (10 mM/100 mM) as shown in Figure S13a. Figure S13b shows the ion mobility for the inorganic (● symbol) and organic cations (▲ symbol) with their chloride counter ions (○△ symbols). With increasing cation radius, the mobility of the hydrated and non-hydrated cations decreased in the similar values. The mobility of a small organic cation ($TMA^+$) with minor water solvation shell was similar to $Na^+$, whereas others ($TEA^+$/$TPrA^+$), which are believed to have no hydration shells,[36] exhibited ion mobility closed to that of inorganic cations of comparable sizes. This indicated that that size exclusion predominantly affected on transport across the membrane rather than the influence of charge density for di- and trivalent ions ($Ba^{2+}$/$Ce^{3+}$).

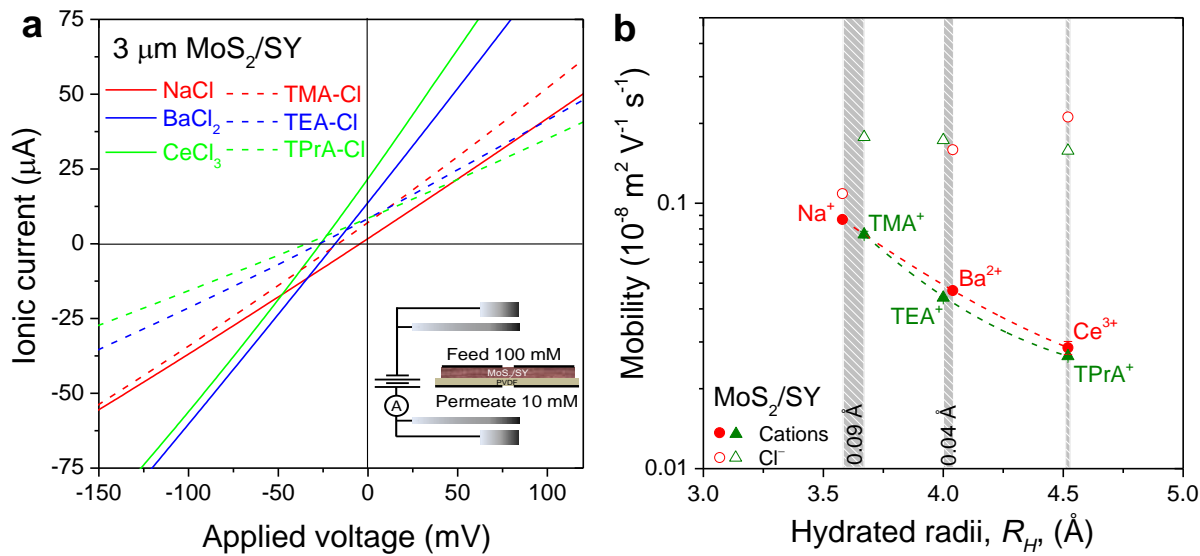

**Figure S13.** Comparison between hydrated and non-hydrated cations. (a) *I–V* characteristics of $MoS_2$/SY (3.01 ± 0.13 μm thick) using chloride solutions with inorganic (solid lines) and organic (dash lines) cations, measured under concentration ratio of 10 (10 mM/100 mM). (b) Ion mobility for inorganic (● symbol) and organic (▲ symbol) cations with similar cation radii as well as their chloride counter ions (○△ symbol), respectively. The different cation sizes are shown in the grey shaded areas with the difference in the hydrated radii (metal ions) and bare radii (organic ions) added. The data shown in (b) is derived from mobility ratio (zero-current potential) at a concentration ratio of 10 and ion conductivity at a constant salt concentration of 100 mM.

## 13. Voltage drop across a bare PVDF membrane

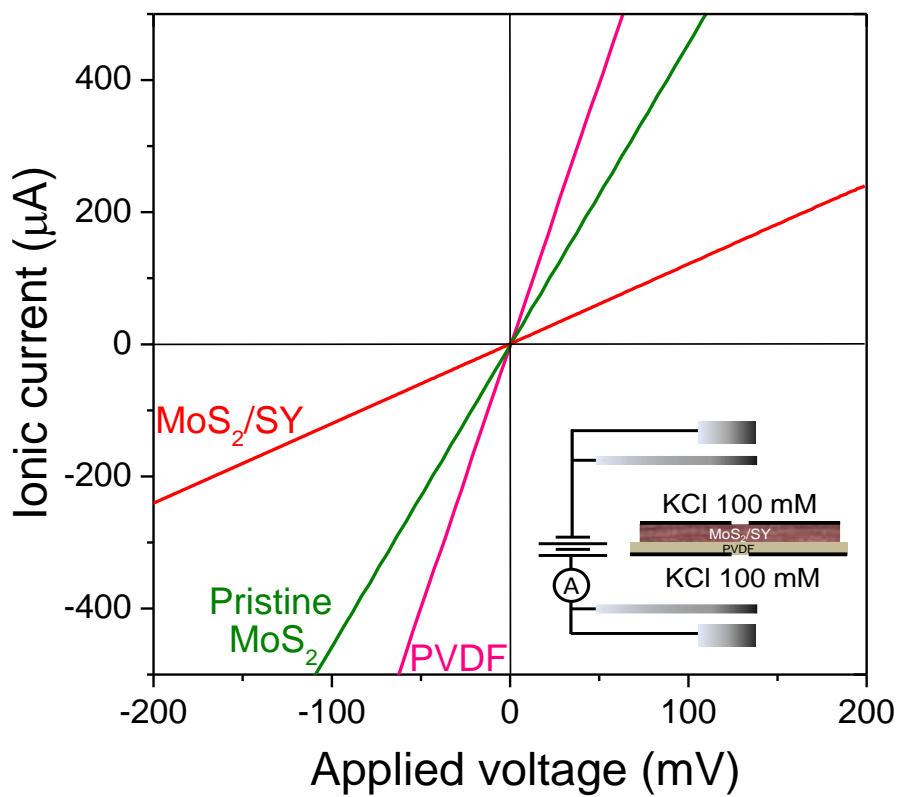

**Figure S13.** *I-V* characteristics of the MoS$_2$/SY membrane (red line), a pristine exfoliated MoS$_2$ membrane (green line) as the comparable thickness (~3 μm thick), and a bare PVDF membrane (pink line) under a constant KCl concentration between feed and permeate reservoirs (100 mM). A bare PVDF membrane contributes 20% or less to the overall resistance of the MoS$_2$ membranes with PVDF as supporting material.

## 14. Supporting References